\documentclass[preprint,aip,preprintnumbers,amsmath,amssymb]{revtex4-1}

\usepackage{graphicx}
\usepackage{dcolumn}
\usepackage{bm}
\usepackage{amsmath}

\begin{document}


\title{Symmetry-dependent transport behavior of graphene double dots}

\author{Paolo Marconcini} \email[Corresponding author: ]
{paolo.marconcini@iet.unipi.it}
\author{Massimo Macucci}
\affiliation{Dipartimento di Ingegneria dell'Informazione, 
Universit\`a di Pisa, Via Caruso 16, I-56122 Pisa, Italy.}

\begin{abstract}
By means of an envelope function analysis, we perform a numerical investigation
of the conductance behavior of a graphene structure consisting of two 
regions (dots) connected to the entrance and exit leads through constrictions 
and separated by a potential barrier.
We show that the conductance of the double dot depends on the symmetry of the 
structure and that this effect survives also in the presence of a low level of 
disorder, in analogy of what we had previously found for a double dot obtained 
in a semiconductor heterostructure.
In graphene, this phenomenon is less dramatic and, in particular, conductance 
is not enhanced by the addition of symmetric constrictions with respect to 
that of the barrier alone.
\end{abstract}

\maketitle

\section{Introduction}
\label{introduction}

In the last decades, the improvement of nanofabrication techniques
has allowed achieving rather large values for the mean free path and the
phase relaxation length
in the high-quality materials used for low-dimensional devices. 
In these conditions, the phase coherence of the electron
wave function is preserved on length scales larger than the device
size, and thus, the wavelike nature of electrons plays a role in the
device behavior.
In particular, the interplay between the phases of different
components of the wave function can lead to impressive effects, such as
resonant tunneling or other interference phenomena.

This is in particular true for semiconductor heterostructures,
where a high-mobility two-dimensional electron gas (2DEG) is present
at the interface between materials with different bandgap but compatible
lattice constant. A significant research activity has focused on ballistic or 
quasi-ballistic devices fabricated at the micro- and nano-scale by confining 
the 2DEG present in such 
heterostructures~\cite{wees,pepper,kastner,smith,heiblum,glattli,mitin,roblin,
ferry,ehrenreich,harrison,epl,prb2009,fnl,cavita}.
We have recently discovered~\cite{jce,prl} an interesting conductance 
enhancement effect that occurs in symmetric mesoscopic cavities with a barrier
in the middle. A mesoscopic cavity is a region of 2-dimensional electron
gas connected to input and exit leads (parallel to the transport direction) 
by constrictions that are usually much narrower than the cavity.
Such a cavity can be defined in semiconductor heterostructures by means
of a combination of etching and negatively biased depletion gates~\cite{ober1}.

We have observed that if a transversal barrier is included (obtained,
for example, with an additional depletion gate), splitting the cavity
into two dots, the conductance of the overall device is
strongly dependent on the symmetry: it reaches a maximum value when 
the barrier is exactly in the middle between the two constrictions. Such an 
effect occurs on a wide energy range; thus, it is not just a resonance.
If the symmetry is broken, for example by shifting the position of the 
barrier or introducing a magnetic field perpendicular to the 2-DEG, the 
conductance quickly drops. 
Furthermore, we noticed also that, in symmetric conditions, the conductance 
can (somewhat counterintuitively) be greater than that of an analogous 
structure with the barrier alone. Thus, the addition of the symmetrically
located constrictions defining the cavity enhances, instead of decreasing,
the conductance.

We found that the origin of the effect is the interference between the
different Feynman's paths into which the electron wave function originating
from the entrance lead can be split. While propagating in the double
dot, each of these paths generally impinges several times against the
barrier, and each time is either transmitted or reflected. In
the case of perfect symmetry, several of these paths, which reciprocally
differ only for exactly symmetrical segments, have the same phase, and thus 
constructively interfere, giving rise to an increase of the overall
conductance with respect to conditions in which the symmetry is
destroyed. 

By means of numerical simulations, we have shown that the dependence of 
conductance on symmetry survives also in the presence of non-idealities 
such as low levels of potential disorder~\cite{jap}. For this reason,
we have suggested the possibility of using such structures to reveal the 
presence of symmetry-breaking factors, i.e., to make position or magnetic
field sensors. 

Here we focus instead on the possibility of observing an analogous effect
in graphene-based double dots.

Graphene is a recently isolated~\cite{geim} two-dimensional
material made up of a honeycomb lattice of carbon atoms, characterized
by quite interesting properties, such as high mobility, thermal conductance,
and mechanical strength. In the last few years, a large theoretical and
experimental research activity has focused on this new material (see, for
example, Refs.~\cite{rise,status,castroneto,experimental,mikhailov,
connolly,iwceconn,acsnano,iwceroche,raza,schwierz,persp,novoselov}), and
a variety of applications have been proposed in different fields, ranging
from electronics, optoelectronics, and sensor fabrication, to thermal and
energy and gas storage applications, to mechanics and membrane production.

As a consequence of its peculiar lattice structure, in monolayer graphene the
envelope functions are governed by the Dirac-Weyl equation~\cite{ando,kp}
(i.e., the same equation which describes the quantum relativistic behavior of
massless fermions), instead of by an effective mass Schr\"odinger equation.
This is at the origin of some peculiar graphene transport
properties, which resemble relativistic phenomena, such as the Zitterbewegung 
and the Klein tunneling~\cite{relativistic,katsnelson,beenakker}. This last 
effect, consisting in an increase of electron tunneling through a tunnel 
barrier,
with unitary transmission in the case of an electron orthogonally impinging
against it (whatever the barrier height), is due to the perfect matching of the
electron states outside the barrier with the hole states inside the barrier,
with conservation of graphene pseudospin. 

The long phase relaxation length of graphene, which allows the conservation
of the phase of the electron wave function over significant lengths, in
principle makes this material a good alternative candidate
for the fabrication of double-dot sensors based on the previously described
symmetry-related effect. The adoption of graphene would make it possible to
decrease the production costs of these sensors and to reduce their thickness,
making them at the same time flexible and transparent, as required in several
applications.

Here we will show, by means of numerical simulations based on the Dirac 
equation, that this effect exists in graphene, too, even in the presence of 
a moderate potential disorder. However, the peculiar transport
properties of graphene make the overall effect less dramatic.

\section{Numerical technique}
\label{technique}

In our study, we consider monolayer graphene ribbons with armchair edges.
Two constrictions, created, for example, by etching, 
separate a part of the ribbon, representing the cavity, from the entrance
and exit leads. The cavity is partitioned into two dots by a potential 
barrier, which, for numerical convenience, we consider to be rectangular. 
It could be obtained by etching and/or with negatively biased gates located 
at a certain distance from the graphene ribbon.
From a practical point of view, the unavoidable asymmetries in the geometry 
of the double dot deriving from the impossibility to control the atomic details
of the device could be overcome by means of a proper calibration procedure,
based on depletion gates, in analogy with what has been proposed in 
Ref.~\cite{jap} for cavities defined in the 2DEG of semiconductor 
heterostructures.

\begin{figure}
\begin{center}
\includegraphics[width=8cm]{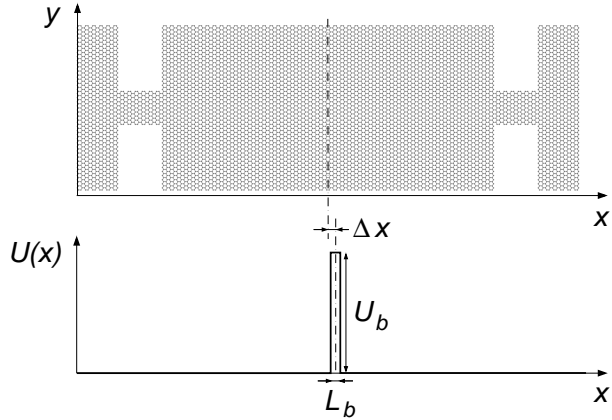}
\caption{Sketch of the graphene double dot.}
\label{figure1}
\end{center}
\end{figure}

A sketch of the simulated structure is shown in Fig.~\ref{figure1}.

The overall wave function of graphene $\psi (\vec{r}\,)$ can be written
as~\cite{kp}
\begin{equation}
\psi (\vec{r}\,)=\sum_{\beta=A,B}\sum_{\vec{R}_{\beta}}
\psi_{\beta} (\vec{R}_{\beta})\varphi(\vec{r} -\vec{R}_{\beta})\, ,
\end{equation}
where $\vec{R}_A$ and $\vec{R}_B$ represent the positions of the carbon atoms
on the two sublattices $A$ and $B$ and $\varphi(\vec{r})$ is the $2 p^z$
atomic orbital of each carbon atom. Using an envelope function approach, the
wave functions $\psi_A (\vec r)$ and $\psi_B (\vec r)$ corresponding to the
two sublattices can be expressed as
\begin{equation}
\begin{cases}
\displaystyle \psi_A (\vec r) = 
e^{i \vec K \cdot \vec r} F^{\vec K}_A (\vec r)
-i\, e^{i \vec K' \cdot \vec r} F^{\vec K'}_A (\vec r)\\[5pt]
\displaystyle \psi_B (\vec r) = 
i\,e^{i \vec K \cdot \vec r} F^{\vec K}_A (\vec r)
+e^{i \vec K' \cdot \vec r} F^{\vec K'}_A (\vec r)\,,
\end{cases}
\end{equation}
where the envelope functions $F^{\vec \alpha}_{\beta} (\vec r)$ (corresponding
to the two inequivalent Dirac points $\vec \alpha=\vec K,\vec K'$ and
sublattices $\beta=A,B$) have to satisfy the Dirac-Weyl equation
\begin{equation}
\begin{cases}
\displaystyle 
[-i\gamma(\partial_x\sigma_x+\partial_y\sigma_y)+U(\vec{r})I]
\vec{F}^{\vec{K}}(\vec r)= E\vec{F}^{\vec{K}}(\vec r)\\[5pt]
\displaystyle
[-i\gamma(\partial_x\sigma_x-\partial_y\sigma_y)+U(\vec{r})I]
\vec{F}^{\vec{K'}}(\vec r)= E\vec{F}^{\vec{K'}}(\vec r)
\end{cases}\!\!
\end{equation}
with boundary conditions depending on the shape of the particular graphene
structure. Here, $\vec{F}^{\vec{K}}=[\vec{F}^{\vec{K}}_A,\ \vec{F}^{\vec{K}}_B]^T$,
$\vec{F}^{\vec{K'}}=[\vec{F}^{\vec{K'}}_A,\ \vec{F}^{\vec{K'}}_B]^T$, the
$\sigma$'s are the Pauli matrices, $E$ is the injection energy, $U$ is the
potential energy, and $\gamma=\hbar v_F$ (with $\hbar$ the reduced Planck
constant and $v_F$ the graphene Fermi velocity). 

In analogy with Ref.~\cite{icnf2013}, we have partitioned the considered
device into transversal sections of constant width, within each of which
the potential energy is approximately constant in the transport direction $x$.
Therefore, each of these sections is an armchair graphene ribbon with the 
potential varying only in the transverse direction $y$ and with its
effective edges (the rows of lattice points immediately outside the ribbon)
at $y=0$ and $y=\tilde W$. Since at these two edges points of both the 
$A$ and the $B$ sublattice are present, along them we have to enforce 
the vanishing of both $\psi_A$ and $\psi_B$. Moreover, exploiting
the translational invariance along $x$, we can write each
envelope function in the form~\cite{kp,brey}
$F_{\beta}^{\vec{\alpha}} (\vec{r})=
e^{i \kappa_x x}\,\Phi_{\beta}^{\vec{\alpha}} (y)$,
where $\kappa_x$ is the wave vector in the transport direction.
Therefore the problem to be solved becomes
\begin{equation}
\begin{cases}
\displaystyle \left(\sigma_x \epsilon(y)-\sigma_z\partial_y\right)
\vec{\varphi}^{\vec K}(y)=\kappa_x \vec{\varphi}^{\vec K}(y)\\[5pt]
\displaystyle \left(\sigma_x \epsilon(y)+\sigma_z\partial_y\right)
\vec{\varphi}^{\vec K'}(y)=\kappa_x \vec{\varphi}^{\vec K'}(y)\\[5pt]
\displaystyle \vec\varphi^{\vec K}(0) =\vec\varphi^{\vec K'}(0)\\[5pt]
\displaystyle \vec\varphi^{\vec K}(\tilde W) =
e^{2 i K \tilde W} \vec\varphi^{\vec K'}(\tilde W)\,,
\end{cases}
\label{sys}
\end{equation}
where $\vec\varphi^{\vec K}=[\Phi^{K}_A,\Phi^{K}_B]^T$,
$\vec\varphi^{\vec K'}=i\,[\Phi^{K'}_A,\Phi^{K'}_B]^T$, and
$\epsilon(y)=(E-U(y))/\gamma$.

We have verified that the solution of this problem in the direct space with
standard discretization techniques yields a large number of spurious solutions
or the appearance of an unphysical degeneracy. The adoption of the Stacey
discretization scheme allows to overcome these
problems~\cite{stacey,tworzydlo} but still represents
an inefficient solution method, because it requires a quite dense 
discretization
grid in order to achieve a satisfactory representation of the derivatives.
Therefore, we have used a different solution technique: we have first
introduced, over a doubled domain $[0,2\tilde W]$, the new functions
$\vec\varphi(y)$, defined as $\vec{\varphi}^{\vec K}(y)$ in $[0,\tilde W]$
and as $e^{i 2 K \tilde W}\,\vec{\varphi}^{\vec K'}(2\tilde W-y)$ in
$[\tilde W,2\tilde W]$, and $g(y)=\epsilon(\tilde W-|\tilde W-y|)$. In terms
of these functions, the system \eqref{sys} becomes
\begin{equation}
\begin{cases}
\displaystyle -\sigma_z \partial_y\vec\varphi(y)+\sigma_x g(y)\vec\varphi(y)=
\kappa_x \vec\varphi(y)\\[5pt]
\displaystyle e^{-i 2 K \tilde W}\vec\varphi(2\tilde W)=\vec\varphi(0)\,,
\end{cases}
\end{equation}
where $K=4\pi/(3a)$ ($a=2.46~\AA$ being the graphene lattice constant) and the
boundary condition is periodic for the function
$e^{-i K y}\vec\varphi(y)$~\cite{iwce2010,pt}.
We have solved this new problem in the Fourier domain (but we could 
equivalently
operate in the direct domain using a basis of sinc-related 
functions~\cite{sinc}).
From the solution of this problem, we have obtained the longitudinal wave 
vectors
and the transverse component of the envelope function in each transverse 
section.

The interfaces between the transversal sections are
potential and/or geometrical discontinuities (each geometrical discontinuity
includes also zigzag transversal segments, which are part of the boundary
of the overall graphene structure).

We have computed the scattering matrix of the region straddling
each discontinuity by enforcing the continuity of
the wave function between the left and right side. In general, this has to
be done both for the $A$ and for the $B$ sublattices (i.e., both for $\psi_A$
and for $\psi_B$). We have projected these continuity equations onto a set of
functions $\sin(n \pi y / \tilde W)$ (with $n$ an integer number),
which represent a basis for the transverse components of the envelope functions
that we have found, obtaining in this way a linear system in the
reflection and transmission coefficients.

The interface between two graphene sections of different width includes also 
sections of zigzag boundary. Along such sections we have to
enforce the vanishing of the wave function only for one sublattice (exactly
as it is necessary to do along the edges of zigzag ribbons~\cite{brey}). This
is obtained simply by projecting the continuity equation for that sublattice
onto the sine functions corresponding to the wider graphene section and the
continuity equation for the other sublattice onto the sine functions
corresponding to the narrower graphene section.

Once we have obtained the scattering matrices of all the regions, we can
recursively compose them, in such a way as to obtain the overall 
scattering matrix and, from it, the transmission matrix of the complete device.
Finally, we compute the conductance from the transmission matrix using
the Landauer-B\"uttiker formula~\cite{buttiker}
\begin{equation}
G=\frac{2\,e^2}{h}\,\sum_{i} t_i\,,
\end{equation}
where $e$ is the modulus of the electron charge, $h$ is Planck's constant, and
the $t_i$ are the eigenvalues of the matrix $t^{\dag}t$ (where $t$ is the
global transmission matrix).

The results that we show in the following have been obtained for zero 
temperature and vanishing bias voltage, i.e., without any average over the
injection energy. For finite temperatures, and room temperature, in particular,
there would be some smoothing, due to energy averaging, and additional, 
inelastic scattering due to phonons. Indeed, for the considered structure 
(to be fabricated with graphene on a boron nitride 
substrate, as it will be discussed in the next section), the scattering
due to impurities included in our model is expected to be of the same order 
of magnitude as that due to phonon scattering at 300~K;~\cite{ferrypap,zomer}
therefore, our results should represent a reasonable estimate also for room
temperature operation. 

\section{Numerical results}
\label{results}

\begin{figure}
\begin{center}
\includegraphics[width=8cm]{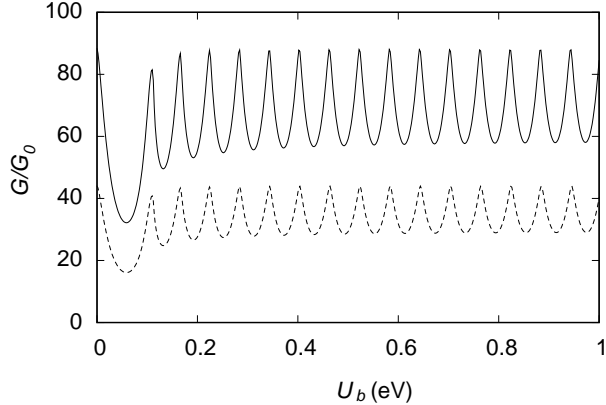}
\caption{Conductance $G$ (normalized with respect to the conductance
quantum $G_0=2\,e^2/h$) obtained for a 2~$\mu$m wide (solid curve) or
1~$\mu$m (dashed curve) wide ribbon with a 30~nm thick barrier, as a
function of the barrier height $U_b$, for an injection energy $E=$40~meV.}
\label{figure2}
\end{center}
\end{figure}

We start by studying the conductance of the barrier alone, without the 
constrictions defining the cavity. 
In Fig.~\ref{figure2}, we report the conductance obtained for a 2~$\mu$m
wide (solid curve) or 1~$\mu$m (dashed curve) wide ribbon with a transverse
barrier (with thickness $L_b=$30~nm) extending across the whole wire width, as
a function of the barrier height $U_b$, for a carrier injection energy equal
to $E=$40~meV.
As expected,~\cite{nguyen} we observe an oscillating behavior, with complete
barrier transparency, when $E\approx U_b-m \gamma \pi/L_b$ (with $m$ an
integer number), which corresponds (neglecting the transverse wave vector)
to the condition of the injection energy being approximately equal to 
the energy of a hole state quasi-confined inside the barrier.

We have then simulated the transport behavior of a 4~$\mu$m long and
2~$\mu$m wide graphene cavity connected to the input and output leads
by two 400~nm wide constrictions and divided into two square
dots by a 30~nm thick and 50~meV high tunnel barrier. More precisely, the wide 
and narrow parts of the cavity are armchair graphene ribbons with 16263 and 
3253 dimer lines across their width, respectively, both corresponding to 
ribbons with a semiconducting behavior.

\begin{figure}
\begin{center}
\includegraphics[width=8cm]{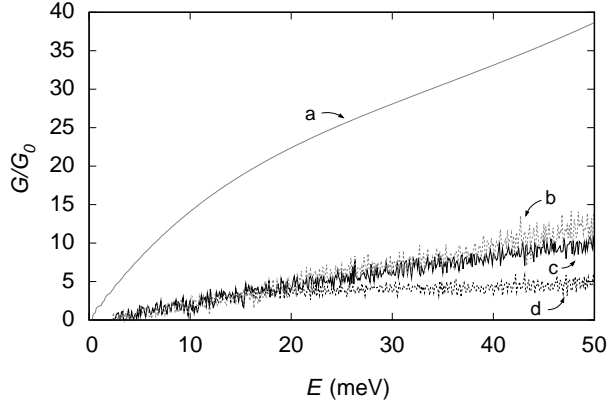}
\caption{Conductance, as a function of the injection energy, of the tunnel
barrier alone (grey solid curve a), of the cavity alone (grey dotted curve b),
of the cavity with the barrier exactly in the middle (black solid curve c),
and of the cavity with the barrier shifted from the center by 10~nm along
the transport direction (black dotted curve d) for a 4~$\mu$m long and
2~$\mu$m (16263 dimer lines) wide graphene cavity, with 400~nm (3253 dimer
lines) wide constrictions and a 30~nm thick and 50~meV high tunnel barrier.}
\label{figure3}
\end{center}
\end{figure}

\begin{figure}
\begin{center}
\includegraphics[width=8cm]{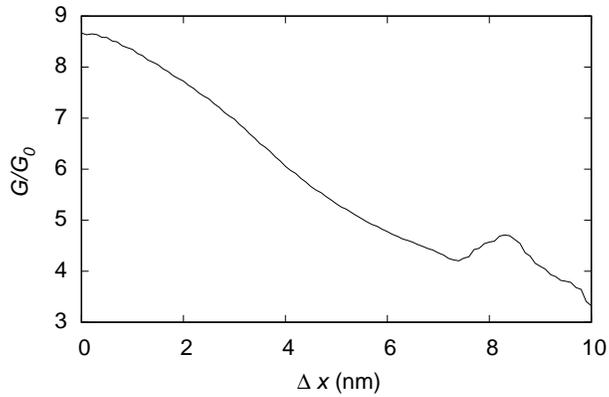}
\caption{Conductance of the double dot considered in Fig.~\ref{figure3}
as a function of the shift $\Delta x$ of the barrier from the center,
for an injection energy of 40~meV.}
\label{figure4}
\end{center}
\end{figure}

In Fig.~\ref{figure3}, we report the conductance, as a function of the injection
energy, of the tunnel barrier alone (grey solid curve a), of the cavity alone
(grey dotted curve b), of the cavity with the barrier exactly in the middle
(black solid curve c), and of the cavity with the barrier shifted from the
center by 10~nm (black dotted curve d).
Comparing these results with those obtained in the case of semiconductor
heterostructures,~\cite{jce,prl,jap} we see that here the presence of two
constrictions symmetrically located around the barrier does not
enhance its transmission, but rather decreases it by a significant amount.
However, if the barrier is shifted away from the center of the cavity 
(i.e., the symmetry is broken), 
the transmission exhibits a clear decrease, even though this effect is now 
apparent
only for higher energies and is less dramatic than in semiconductor
heterostructures.

Indeed, as we observed in Refs.~\cite{prl} and \cite{jap},
the dependence of the conductance of a double dot on symmetry increases as
the barrier is made less transparent. Since in graphene the transparency of
the barrier is enhanced by Klein tunneling, the dependence of conductance
on symmetry is weaker than in semiconductor heterostructures. 
Also the fact that conductance is never increased over that of the barrier
alone (as it is instead observed in analogous devices implemented with III-V
material systems) could be explained by
the presence of Klein tunneling, which significantly increases the 
transmission for electrons, in particular those that impinge orthogonally 
against the barrier. 

In Fig.~\ref{figure4} we show the dependence of the conductance of the double
dot on the shift $\Delta x$ of the barrier from the center, keeping the
injection energy constant at 40~meV. As we see, the conductance has a maximum
when the barrier is exactly at the center of the cavity, while it decreases as 
the barrier is shifted away. The presence of a smaller local
maximum for a barrier shift of about 8.3~nm can be attributed to the
constructive interference of a small subset of transmission paths for that
barrier position.

Analogous simulations have been performed also for a very similar
structure, in which, however, the wide and narrow parts of the device 
are 16262 and 3254 dimer line wide, respectively, and thus are characterized
by a metallic behavior.
The results we have obtained are very similar to those shown in
Figs.~\ref{figure3} and \ref{figure4}. This demonstrates that the
phenomenon does not depend on the exact number of atoms across the width
of the wire, and in particular on the semiconducting or metallic nature of
the graphene sections.

We have investigated how the observed effect depends on cavity size, by
performing simulations for smaller cavities. Here we report a couple of 
meaningful examples: a $2 \times 1$~$\mu$m$^2$ cavity and a 
$400 \times 200$~nm$^2$ cavity. It would be interesting to reduce the 
size of the cavity to have less demanding requirements in terms of mean 
free path, but, on the other hand, the magnitude of the conductance modulation
is reduced, too, in smaller cavities.  

\begin{figure}
\begin{center}
\includegraphics[width=8cm]{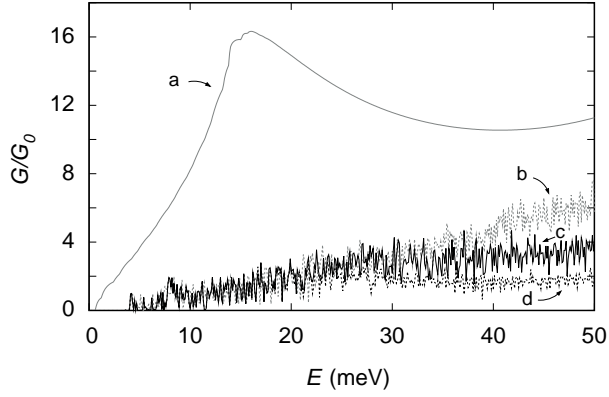}
\caption{Conductance, as a function of the injection energy, of the tunnel
barrier alone (grey solid curve a), of the cavity alone (grey dotted curve b),
of the cavity with the barrier exactly in the middle (black solid curve c),
and of the cavity with the barrier shifted from the center by 10~nm
(black dotted curve d) for a 2~$\mu$m long and
1~$\mu$m (8131 dimer lines) wide graphene cavity, with 200~nm (1627 dimer
lines) wide constrictions and a 55~nm thick and 50~meV high tunnel barrier.}
\label{figure5}
\end{center}
\end{figure}

\begin{figure}
\begin{center}
\includegraphics[width=8cm]{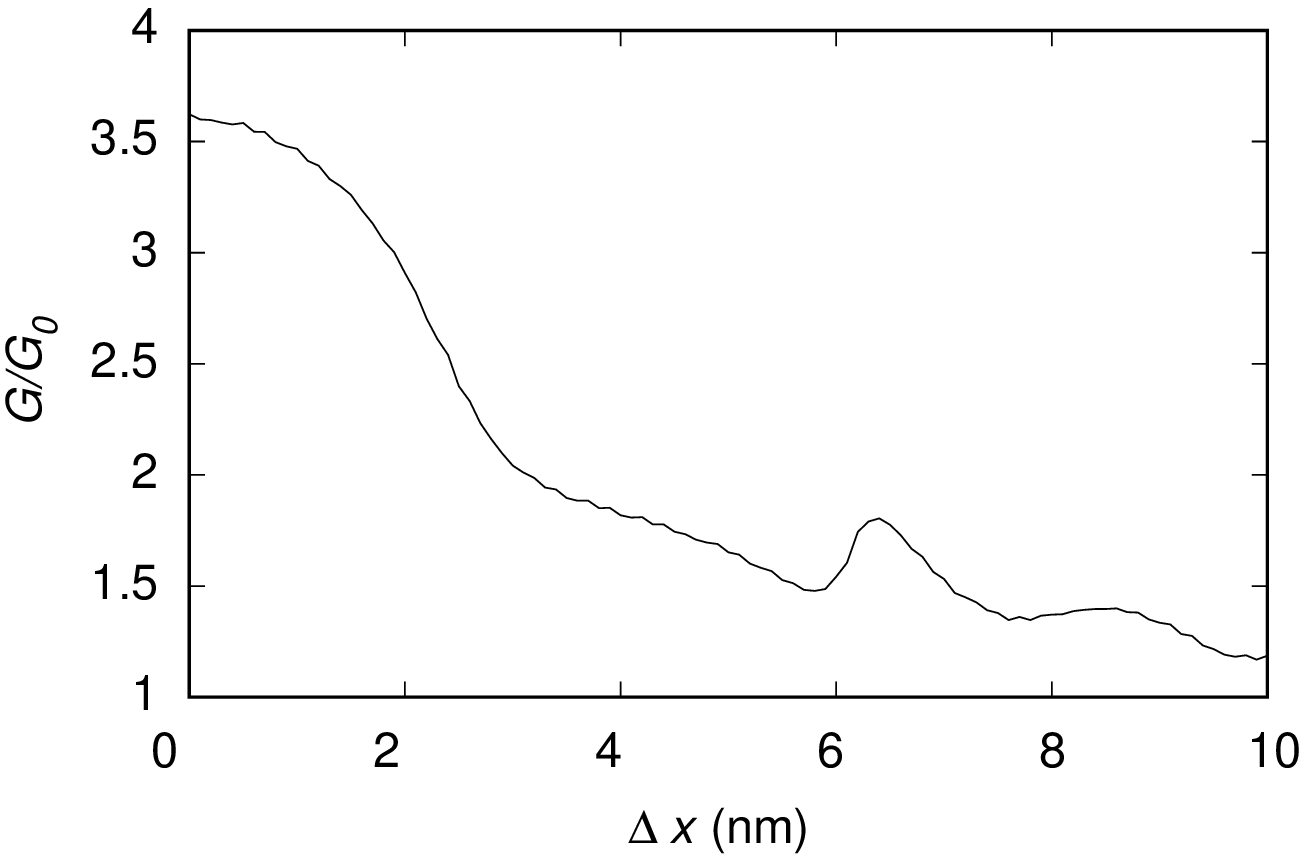}
\caption{Conductance of the double dot considered in Fig.~\ref{figure5}
as a function of the shift $\Delta x$ of the barrier from the center,
for an injection energy of 40~meV.}
\label{figure6}
\end{center}
\end{figure}

For the $2$~$\mu$m long and $1$~$\mu$m wide cavity, with
200~nm wide constrictions and a $U_b=50$~meV high and $L_b=55$~nm
thick tunnel barrier, in Fig.~\ref{figure5} we report the conductance
of the barrier and of the cavity alone, as well as of the cavity 
with the barrier exactly in the middle and shifted 10~nm away from the 
center.
Notice that the transmission of the barrier has a maximum when the injection
energy is approximately
$U_b-\gamma \pi/L_b=17.15$~meV (i.e., when the energy of the impinging
electron states corresponds to that of the hole states inside the barrier).
In Fig.~\ref{figure6}, we report the conductance of this structure as
a function of the shift of the barrier position from the center of the cavity,
which exhibits a behavior substantially analogous to that of the larger cavity
we previously investigated.

\begin{figure}
\begin{center}
\includegraphics[width=8cm]{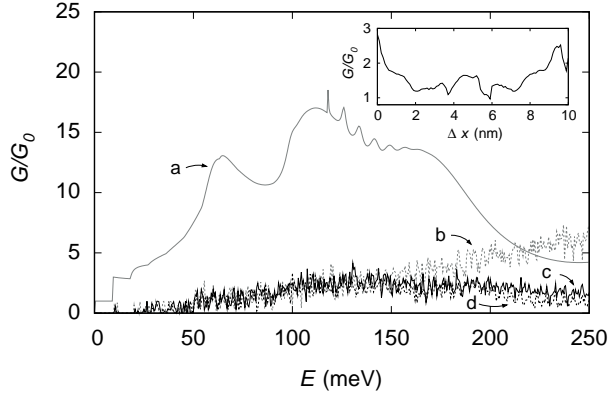}
\caption{Conductance, as a function of the injection energy, of the tunnel
barrier alone (grey solid curve a), of the cavity alone (grey dotted curve b),
of the cavity with the barrier exactly in the middle (black solid curve c),
and of the cavity with the barrier shifted from the center by 10~nm
(black dotted curve d) for a 400~nm long and 200~nm (1625 dimer line) wide
cavity, with 40~nm (323 dimer line) wide constrictions and a 30~nm thick
and 250~meV high tunnel barrier. In the inset we report the conductance
of the double dot as a function of the shift $\Delta x$ of the barrier
from the center, for an injection energy of 200~meV.}
\label{figure7}
\end{center}
\end{figure}

For the 400~nm long and 200~nm
wide cavity, we have considered 40~nm
wide constrictions and a 30~nm thick 
and 250~meV high tunnel barrier. In this case, we have increased the range 
of considered injection energies (and, as a consequence, the height of the 
barrier) in order to have a sufficiently large number of propagating modes
(the effect we are interested in can be observed only in the presence of 
a relatively large number of modes, also in the previously studied cavities).
We report results for the 400~nm $\times$ 200~nm cavity in Fig.~\ref{figure7}: 
in this case, the conductance does not exhibit a significant dependence on the
symmetry of the structure for any injection energy.
This is clear also from the inset of Fig.~\ref{figure7}, where we show the
conductance as a function of barrier shift, for an injection energy of 
200~meV: the conductance decreases monotonically only in a reduced interval
near the origin. 

This confirms a trend that we had already observed for heterostructure-based
cavities~\cite{jap}: there is an optimal cavity size, because on the one hand 
it is better to have a large cavity, in order to enhance the effect, on the 
other hand too large a cavity leads to long trajectories, which involve too 
strong inelastic scattering.

\begin{figure}
\begin{center}
\includegraphics[width=8cm]{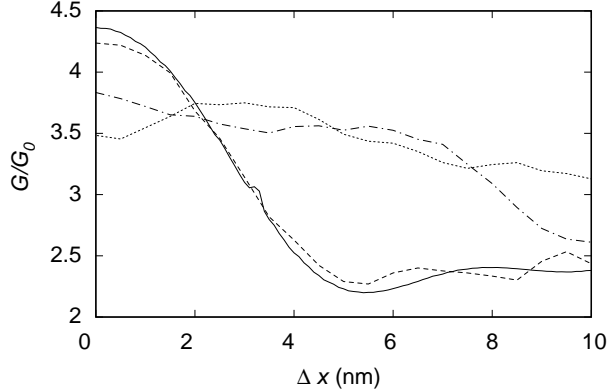}
\caption{Conductance of a 2~$\mu$m long and 1~$\mu$m (8132 dimer lines)
wide graphene cavity, with 200~nm (1626 dimer lines) wide constrictions
and a 30~nm thick and 50~meV high tunnel barrier as a function of the shift
$\Delta x$ of the barrier, for an injection energy of
40~meV, in the absence of disorder (solid curve) and in the presence of a
disorder represented by a distribution (with a concentration equal to
$5\times10^{10}$~cm$^{-2}$) of Gaussian functions with a half-width at
half-maximum equal to 5~nm and an amplitude randomly distributed between
$-10$~meV and $10$~meV (dashed curve), $-25$~meV and $25$~meV (dotted-dashed
curve) and $-50$~meV and $50$~meV (dotted curve).}
\label{figure8}
\end{center}
\end{figure}

Finally, we have studied the influence of potential disorder on the observed
effect. In particular, here we show the results that we have obtained for
a 2~$\mu$m long and 1~$\mu$m wide
cavity with 200~nm wide 
constrictions and a 30~nm thick and 50~meV high tunnel barrier.
To the potential within the cavity, we have added disorder in the form of a
sum of randomly distributed Gaussian contributions (each with a half-width 
at half-maximum equal to 5~nm) with a density
equal to $5\times10^{10}$~cm$^{-2}$.
In Fig.~\ref{figure8}, we report the conductance as a function of the
barrier shift, both in the absence of disorder
(solid curve) and in the presence of potential fluctuations with different
amplitude ranges. The amplitude of the Gaussians is uniformly distributed
within each considered range. The weakest disorder we have considered is
within the range $[-10\,\,{\rm meV}, +10\,\,{\rm meV}]$, which is narrow,
but still realistic, corresponding to that reported for graphene on a 
hexagonal boron nitride substrate~\cite{xue,morgenstern}. Results for 
this disorder level are reported with a dashed curve in Fig.~\ref{figure8}: 
the main conductance modulation effect does survive in this condition.
The effect is instead suppressed if we consider wider amplitude ranges, e.g.,
$[-25\,\,{\rm meV}, +25\,\,{\rm meV}]$ (dashed-dotted line) and 
$[-50\,\,{\rm meV}, +50\,\,{\rm meV}]$ (dotted line), which are closer to
the values measured on samples of graphene on SiO$_2$ 
substrates~\cite{droscher,morgenstern}.

\section{Conclusion}
\label{conclusion}

Our numerical results have shown that the conductance of a graphene double dot
strongly depends on the symmetry of the structure, exhibiting a maximum when 
the tunnel barrier separating the two halves is exactly in the middle. This 
effect can be observed for cavities with a size in the few-micron range, 
does not depend strongly on the width of the dots or on the thickness of the 
barrier, and, in addition, survives in the presence of a moderate level of 
disorder.
Overall, in graphene this phenomenon is less dramatic than in 
heterostructure-based double dots and, in particular, it does not include
the enhancement of the conductance over that of the barrier 
alone~\cite{prl}.

Our analysis in terms of disorder amplitude reveals that for an experimental 
verification of this effect in graphene it will be necessary to use material
of very high quality, deposited on substrates, such as
boron nitride, for which potential fluctuations of the order of 10~meV or less
have been estimated.

Due to the fact that for such material room temperature phonon scattering 
(which is not included in our calculation) is expected to be of the same order 
of magnitude as impurity scattering, there is hope for a working room
temperature demonstration of the effect, and therefore for the practical
implementation of position or magnetic field sensors, which would be unlikely
with III-V semiconductors, whose mobility is degraded too much by the phonon
contribution.

\bibliographystyle{apsrev}

\end{document}